# Polarity induced electronic and atomic reconstruction at NdNiO$_2$/SrTiO$_3$ interfaces

*Ri He,*[1,2,5] *Peiheng Jiang,*[2] *Yi Lu,*[3] *Yidao Song,*[2] *Mingxing Chen,*[4] *Mingliang Jin,*[5] *Lingling Shui,*[1]* *Zhicheng Zhong*[2,6]*

[1]*School of Information and Optoelectronic Science and Engineering, South China Normal University, Guangzhou 510006, China.*
[2]*Key Laboratory of Magnetic Materials Devices & Zhejiang Province Key Laboratory of Magnetic Materials and Application Technology, Ningbo Institute of Materials Technology and Engineering, Chinese Academy of Sciences, Ningbo 315201, China.*
[3]*Institute for Theoretical Physics, University of Heidelberg, Heidelberg 69120, Germany.*
[4]*School of Physics and Electronics, Hunan Normal University, Key Laboratory for Matter Microstructure and Function of Hunan Province, Key Laboratory of Low-Dimensional Quantum Structures and Quantum Control of Ministry of Education, Changsha 410081, China.*
[5]*International Academy of Optoelectronics at Zhaoqing, South China Normal University, Zhaoqing 526238, China.*
[6]*China Center of Materials Science and Optoelectronics Engineering, University of Chinese Academy of Sciences, Beijing 100049, China.*

*Corresponding authors:
E-mail address: Zhicheng Zhong (zhong@nimte.ac.cn); Lingling Shui (shuill@m.scnu.edu.cn)

## *Abstract*

Superconductivity has recently been observed in Sr-doped NdNiO$_2$ films grown on SrTiO$_3$. Whether it is caused by or related to the interface remains an open question. To address this issue, we use density functional theory calculation and charge transfer self-consistent model to study the effects of polar discontinuity on the electronic and atomic reconstruction at the NdNiO$_2$/SrTiO$_3$ interface. We find that sharp interface with pure electronic reconstruction only is energetically unfavorable, and atomic reconstruction is unavoidable. We further propose a possible interface configuration that contain residual apical oxygen. These oxygen atoms lead to hybrids of $d_{z^2}$ and $d_{x^2-y^2}$ states at the Fermi level, which weaken the single-band feature and may be detrimental to superconductivity.

## I. INTRODUCTION

The recent discovery of superconductivity in Sr-doped infinite-layer $NdNiO_2$ films is an important breakthrough in high-temperature superconductors[1], and has generated enormous excitement[2-29]. In experiment, $Nd_{0.8}Sr_{0.2}NiO_2$ films were fabricated by topotactic reduction (with $CaH_2$ as reductant) from perovskite $Nd_{0.8}Sr_{0.2}NiO_3$ films grown on atomically flat $TiO_2$-terminated $SrTiO_3$ (001) substrates. The $Nd_{0.8}Sr_{0.2}NiO_2$ films exhibit a superconducting transition temperature of 9~15 K. The superconductivity in nickelates is seemingly derived from the single-band physics of the Ni $3d_{x^2-y^2}$ orbital[2-10], analogous to the celebrated copper oxides[30-32]. Considering no superconductivity was observed for bulk $Nd_{0.8}Sr_{0.2}NiO_2$[11,12], the possibility of interface superconductivity was also proposed by several studies[11-14]. Indeed, such possibilities are not unprecedented in unconventional superconducting systems. In $FeSe/SrTiO_3$(STO), for example, the interface plays an important role in enhancing phonon coupling in the FeSe [33,34]. Another more closely related example is the $LaAlO_3/SrTiO_3$ (LAO/STO) system[35-37], where the so-called polar discontinuity[38,39] gives rise to a conducting interface that hosts superconductivity. Could it be the case for the nickelates as well? To answer this question, a careful and systematic analysis of the atomic and electronic structure at the $NdNiO_2$/STO interface is indispensable.

The term polar discontinuity refers to a junction between polar and nonpolar layers in oxide heterostructures[38,39]. This concept was most commonly used to explain the conducting behavior at the LAO/STO interface arising from atomic reconstruction [40-45]. In $NdNiO_2$/STO film, simply counting the charge of atomic layers could lead to a similar conclusion. The alternate stacking of positive charged $Nd^{3+}$ and negative charged $NiO_2^{3-}$ layers in $NdNiO_2$ are grown on non-polar STO (001) substrate with charge neutral $SrO^0$ and $TiO_2^0$ layers (Fig. 1), which induces polar discontinuity at the interface. Polar discontinuity would give rise to a huge effective built-in electric field, leading to a divergence of the electrostatic potential with increasing thickness of NdNiO$_2$ film as illustrated in Fig. 1, which is called polar instability[41]. To avoid polar instability, the electronic and atomic reconstruction will occur at the polar discontinuous interface[38]. Electronic reconstruction refers to electronic structure change such as transfer of electrons[38,46], and atomic reconstruction refers to stoichiometry change such as creation of point defects[41,42]. The electronic and atomic reconstructions often determine the properties of the oxide interface[46,47]. The questions following are how these reconstructions can be induced by polar discontinuity, and how do they affect superconductivity at NdNiO$_2$/STO interface.

Here, the electronic and atomic reconstructions in NdNiO$_2$/STO interface were investigated based on density functional theory (DFT) calculation, and charge transfer self-consistent model combined with point-charge lattice model. We show that instead of maintaining a sharp interface, the system is more in favor of a combination of both electronic and atomic reconstructions at the interface to avoid polar instability. It implies that the Nd/TiO$_2$/SrO interface should be rough and contain extra atoms such as apical oxygen. The residual apical oxygen atoms significantly alter the band structure at the interface and likely renders it irrelevant for superconductivity.

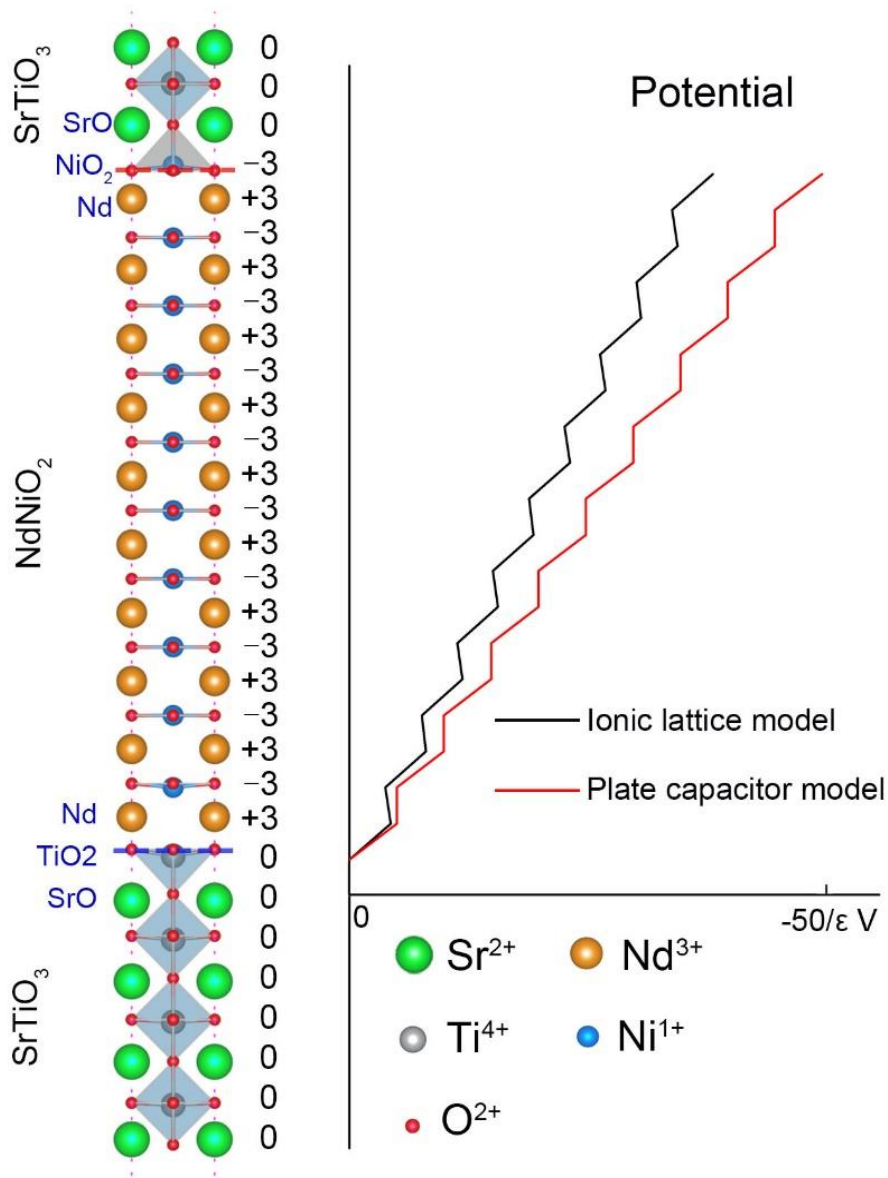

Fig. 1. The polar discontinuity illustrated for atomically sharp interfaces between $NdNiO_2$ and $SrTiO_3$. The unreconstructed interface has neutral (001) planes in $SrTiO_3$, but the (001) planes in $NdNiO_2$ have alternating charges (+3 and −3). It results in an electrostatic potential that diverges with thickness.

## II. METHOD

*DFT calculation.* We studied $NdNiO_2$ thin films in two forms: a $NdNiO_2$ film grown on an STO (001) substrate with a STO capping layer ($(STO)_4|(NdNiO_2)_{10}|(STO)_1$ sandwich structure) and freestanding $NdNiO_2$ films with thickness varying from 1 to 10 unit cells. The vacuum region spans 40 Å to prevent coupling between periodic images.

The 1×1 lattice in *xy*-plane was adopted. The in-plane lattice constants of NdNiO$_2$ were fixed at 3.851 Å for the freestanding ones, and at 3.905 Å for the sandwich structure, which correspond to the calculated value of bulk NdNiO$_2$ and STO, respectively. All internal atomic positions were allowed to fully relax. For the sandwich structure with apical oxygen, an oxygen atom was added in the interfacial Nd layer. The DFT calculations were performed using a plane-wave basis set with a cutoff energy of 500 eV as implemented in the Vienna *ab initio* simulation package (VASP)[48,49], and electron exchange-correlation potential was described using generalized gradient approximation (GGA) and Perdew-Burke-Ernzerhof solid (PBEsol) scheme[50]. The Brillouin zone was sampled with a 12 × 12 × 1 Monkhorst-Pack k-point grid. The magnetism in the nickelate films makes a negligible contribution to total energy compared to electrostatic energy caused by polar discontinuity, and a Coulomb repulsion term U only modifies the electrostatic potential profile slightly and does not change our main conclusions.

*Point-charge lattice model.* The electrostatic potential of alternate stacking layers in NdNiO$_2$ were calculated by a point-charge lattice model. The electrostatic potential $\phi_i$, for atom site *i* is computed as $\phi_i = \sum \frac{Z_j}{4\pi\varepsilon_0\varepsilon r_{ij}}$, where $Z_j$ is the valence (in the unit of the elementary charge) of the *j*th ion with formal valences Ni$^+$, Nd$^{3+}$, or O$^{2-}$. $\varepsilon_0$ is the vacuum permittivity, $\varepsilon$ is the dielectric constant of materials, and $r_{ij}$ is the distance between ions *i* and *j*. The electrostatic potential energy of a crystal per unit cell is calculated as $E_p = \frac{1}{2}\sum(\phi_i Z_i)$ using Evjen method[51]. We found that the atomic relaxation could only slightly reduce the electrostatic potential and electrostatic energy compared to the contribution of polar discontinuity, so we used the calculated bulk NdNiO$_2$ lattice constant (*a* = 3.851 Å, *c* = 3.262 Å) and neglected the atomic relaxation induced by the interface. A plate-capacitor model was also applied to estimate the electrostatic potential for comparison.

*Charge transfer self-consistent model.* For the polar discontinuity LAO/STO interface, it is absent of charge transfer because of the insulating LAO layer[35]. In contrast, NdNiO$_2$ is metallic (Fig. 2)[20], so that the Ni valence can change to partially screen the polar instability. The distribution of Ni valence changes ($\delta Z_{Ni}$) in freestanding NdNiO$_2$ films can be determined by a simple charge transfer self-consistent model, being similar with previous work[52]. We simplified the analysis by assuming the density of states of Ni $d$ orbitals near Fermi level $D$ as a constant, and adopted rigid band approximation. The Ni valence change $\delta Z_{Ni}^i$ (from the nominal value of 1+) in layer $i$ is then realized by shifting the local Fermi energy level by $\delta Z_{Ni}^i/D$ within our approximation. The built-in electric field resulted from electronic reconstruction can be fully or partly compensated and we suppose a residual electric potential of layers $i$ is $\phi_i$. Hence an equilibrium equation of $\delta Z_{Ni}^i/D + \phi_i = 0$ can be obtained. Solving this equation numerically based on DFT calculations, we can obtain the distribution of $\delta Z_{Ni}^i$ to describe the electronic reconstruction.

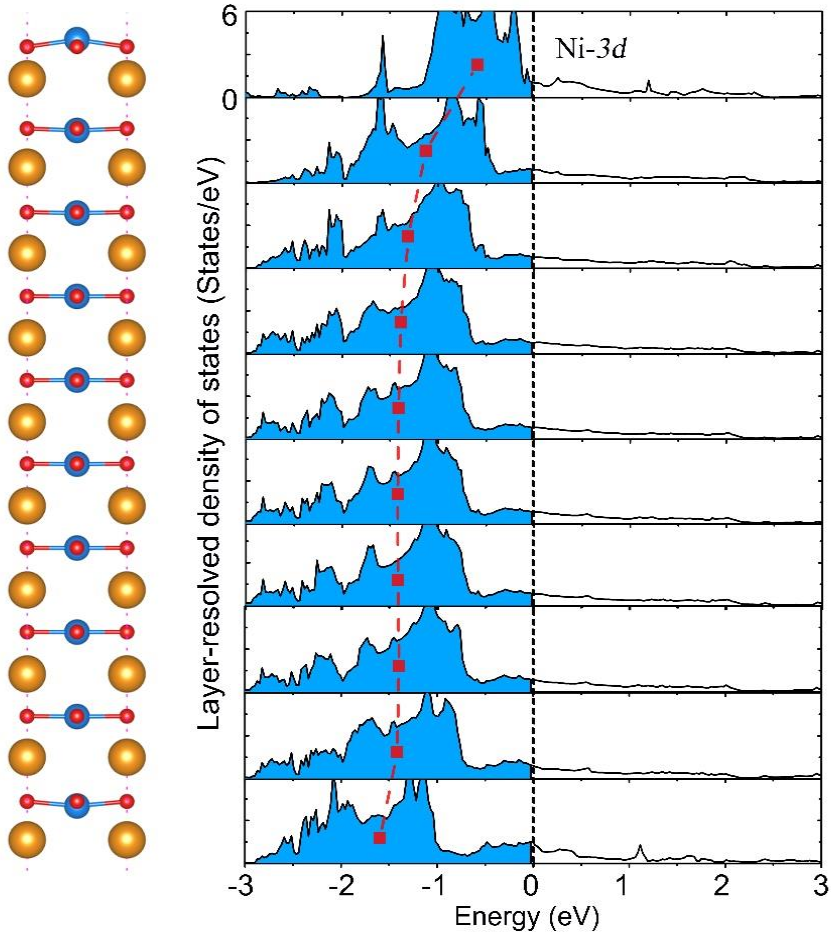

Fig. 2. The optimized atomic structure of freestanding NdNiO$_2$ and corresponding layer-resolved

partial density of states (PDOS) projected on Ni-3$d$ orbitals. The red squares represent weighted average energy in each layers and red dashed line represents the shift of PDOS induced by a huge effective built-in electric field.

## III. RESULTS

*Polar discontinuity induces polar instability.* NdNiO$_2$ films grown on a TiO$_2$-terminated SrTiO$_3$ (001) substrates can be viewed as alternate staking of positively charged Nd$^{3+}$ layers and negatively charged NiO$_2$$^{3-}$ layers. In the absence of electronic reconstruction (i.e. $\delta Z_{Ni}^i = 0$), the polar discontinuity at the interface will lead to a built-in electric field, and thus inducing a huge electrostatic potential. As shown in Fig. 1, the electrostatic potential decreases monotonically across the system by an amount of $\delta V = -5.13/\varepsilon$ V (plate capacitor model) and $-3.95/\varepsilon$ V (point-charge lattice model) per double layer. The electrostatic energy difference between electronic unreconstructed freestanding NdNiO$_2$ and bulk one ($\Delta E(N) = E^{\text{freestanding}}(N) - N \times E^{\text{bulk}}$, where $N$ is the thickness of freestanding film and $E^{\text{bulk}}$ is electrostatic energy of bulk unit cell) is the extra electrostatic energy that induced by polar discontinuity. As shown by the red line in Fig. 3, without considering electronic reconstruction, $\Delta E$ increases linearly with thickness $N$, and extrapolates to $\Delta E = 44.2 \times N/\varepsilon$ eV. It can be seen that $\Delta E$ is so large, even for ultrathin films, that it will lead to structure instability of the system[39,41,52].

*Electronic reconstruction.* Considering the metallicity of NdNiO$_2$ (Fig. 2)[20], electronic reconstruction is therefore expected to happen to screen the huge effective built-in electric field and reduces the electrostatic energy. Here, we performed both DFT calculations and lattice model analysis to investigate the electronic reconstruction. We note that for NdNiO$_2$/STO, the polar discontinuous interface is the same as the surface of freestanding NdNiO$_2$ film due to the neutral layers in STO. The optimized structure of 10-unit-cell freestanding NdNiO$_2$ film and corresponding layer-resolved partial density of states (PDOS) of Ni-3$d$ orbitals are shown in Fig. 2. It indicates that an effective built-in electric field induces a substantial charge transfer from top surface Ni to bottom. The weighted average energy of PDOS relative to Fermi level is shifted in surface layers, but relaxes to an almost constant value when moved away from the intermediate interface. This shows that the built-in electric field is fully compensated in central bulk region by charge transfer, and the electrostatic energy of the system is dramatically reduced.

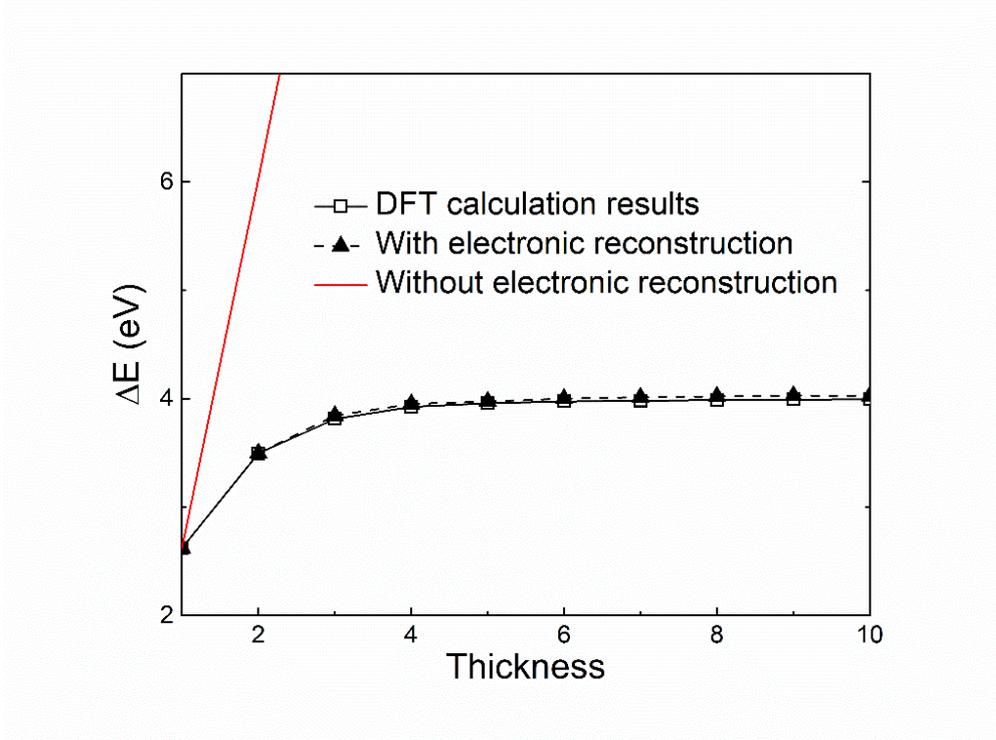

Fig. 3. Total energy difference between the freestanding and bulk NdNiO$_2$ as a function of the thickness. The red line and black triangles represent the results of point-charge model analysis without and with consideration of electronic reconstruction for freestanding NdNiO$_2$, respectively. The open squares are DFT calculated results.

To quantitatively estimate the reduction of electrostatic energy due to the electronic reconstruction, we calculate the total energy difference between the electronic reconstructed freestanding system and bulk one, $\Delta E(N) = E^{\text{freestanding}}(N) - N \times E^{\text{bulk}}$ as a function of thickness $N$ by DFT calculation. This $\Delta E$ can be regarded as electrostatic energy that induced by polar discontinuity after electronic reconstruction. As shown in Fig. 3, $\Delta E$ converges with increasing thickness $N$ for freestanding NdNiO$_2$ to ~4.03 eV above a critical thickness of 6 unit cells, in stark comparison with the linearly diverging behavior of the unreconstructed case. The DFT result can be faithfully reproduced by the point-charge lattice model using parameter values $D = 0.59$ e/V, $\varepsilon = 10$, and $E_s = 1.798$ eV. The surface energy $E_s$ is added as an additional parameter here as it is not intrinsically included in the point-charge lattice model.

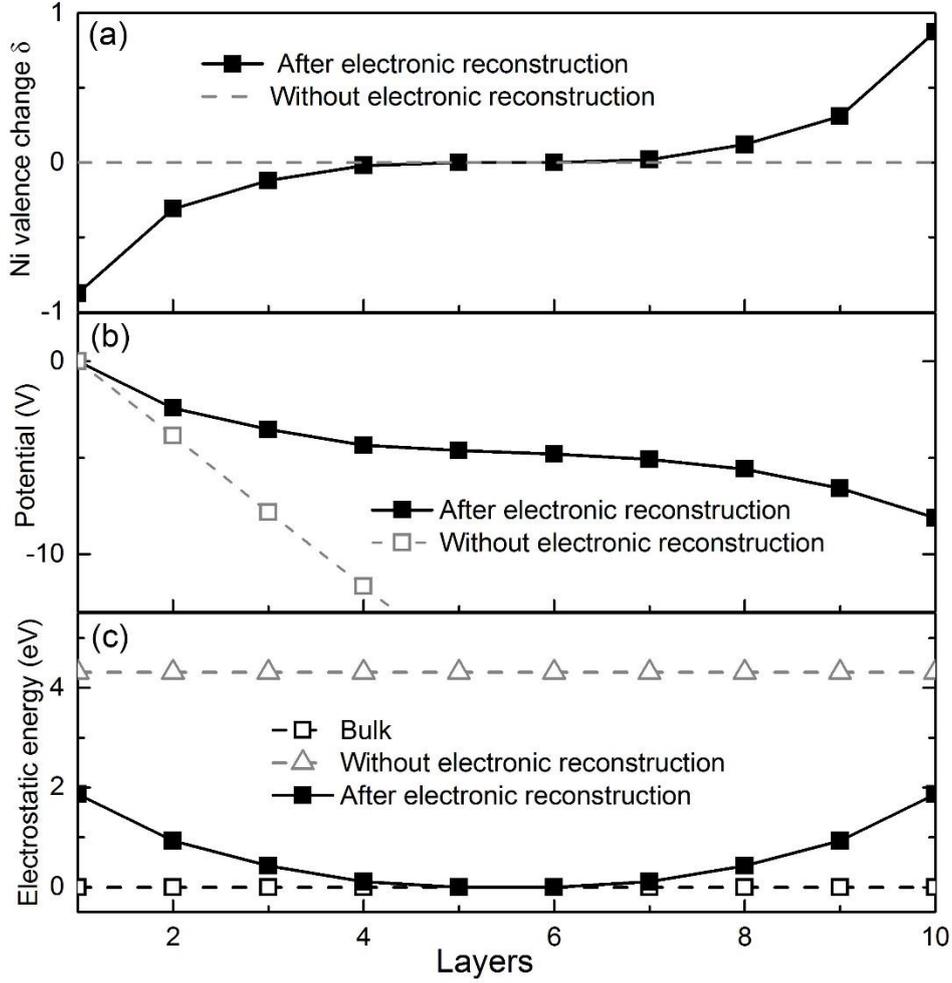

Fig. 4. (a) Position dependence of Ni valence change δ based on the charge transfer self-consistent model including density of states described in the main text. (b) The electrostatic potential of multilayers before and after electronic reconstruction. (c) The layer-resolved electrostatic energy with respect to bulk before and after electronic reconstruction.

We use a 10-unit-cell freestanding $NdNiO_2$ films as an example to describe the specific details of electronic reconstruction. Solving the equilibrium equation of $\delta Z_{Ni}^{i}/D + \phi_i = 0$ combines with DFT calculated $\Delta E$ ($N = 10$), we can obtain $\delta Z_{Ni}^{i}$ ($i = 1-10$) as shown in Fig. 4(a). It can be clearly seen that $\delta Z_{Ni}^{i}$ at surface is the largest, and then decays rapidly to zero in fully screened bulk region. The valence change in Ni indicates that electrons are driven by the built-in electric field from the top surface to the bottom interface, which is consistent with the charge transfer described by PDOS in Fig. 2. We calculated the layer dependent electrostatic potential with and without including of electronic reconstruction, as shown in Figs. 4(b). It can be seen that the potential for the electronic reconstruction case almost flatten in bulk region of layer 4 to 7, which further evidences that the built-in electric field is compensated in bulk region. We also calculated layer-resolved electrostatic energy relative to bulk values with and without electronic reconstruction as shown in Fig. 4(c). The electrostatic energy of electronic reconstruction case is much smaller than that of electronic unreconstructed one and almost close to bulk values in bulk region of layer 4 to 7. However, there is a substantial residual electrostatic energy remaining at the surface region compared with corresponding bulk system. As a conclusion, although electronic reconstruction can alleviate the electrostatic instability, it is not enough to fully screen the built-in electric field at the surface, which may further lead to an atomic reconstruction at the interface.

*Atomic reconstruction.* According to the electrostatics analysis of polar discontinuity surface, atomic reconstruction should provide a 1.5 excess $e^-$ (half of original) per unit cell in the surface layer region to fully compensate the built-in electric field[39]. For Nd/TiO$_2$/SrO interface, if we assume that it is absent of electronic reconstitution and have extra $n$ atoms of negative charge $-\sigma$, we can use the formula $(+1)^{Ni} + 2\times (-2)^{O} + (+3)^{Nd} + n\times(-\sigma)^{\text{extra atom}} = -1.5$ to determine the interface configurations. There are many possible rough interface configurations with different type of extra atoms (such as residual oxygen[13] and adsorbent hydrogen[26]) and strontium vacancies[53,54] to provide excess electrons. The most possible situation is that during fabricating NdNiO$_2$ from NdNiO$_3$, some oxygen atoms in the NdO layers have not been properly reduced. Therefore, we assumed that there are 0.75 residual oxygen atoms of $2e^-$ in the Nd layer of interface region per unit cell, which coincidentally gives 1.5 excess $e^-$ to satisfy the formula. We calculated the formation energy with the apical oxygen locates at Nd/TiO$_2$/SrO interface and bulk NdNiO$_2$. The result shows that the formation energy in the Nd layer of Nd/TiO$_2$/SrO interface is not only ~1.1 eV lower than the subsequent layer, but also notable that it is ~1.4 eV lower than that of the bulk structure, indicating that it is much easier for apical oxygen to form at the Nd/TiO$_2$/SrO interface rather than bulk. We performed calculations of electrostatic energy difference between atomic reconstructed 10-unit-cell freestanding film and corresponding bulk structure contain 0.75 apical oxygen for the lattice model with a result of $\Delta E_O$ = 0.10 eV. This is much less than $\Delta E$ = 4.03 eV of the system without apical oxygen (i.e. with only electronic reconstitution). It implies that the difference of formation energy for Nd/TiO$_2$/SrO interface and bulk is mainly due to the contribution by the electrostatic energy. Therefore, it suggests that atomic reconstruction is energetically more favorable.

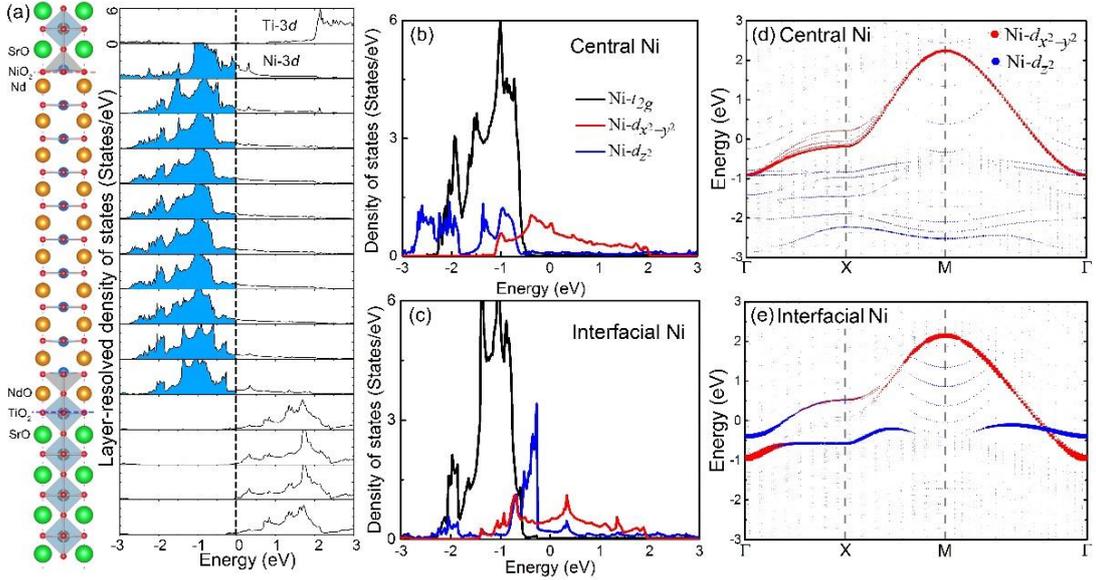

Fig. 5. (a) The optimized atomic structure of freestanding (STO)$_4$/(NdNiO$_2$)$_{10}$/(STO)$_1$ film and corresponding layer-resolved PDOS of Ni-$t_{2g}$, $d_{z2}$ and $d_{x2-y2}$ orbitals. (b, c) Orbital-decomposed PDOS of Ni in the central layer and interfacial layer. (d, e) Orbital projected electronic band structure of central bulk-like Ni and interfacial Ni, the size of the red and blue dots represents the weight of Ni-$d_{x2-y2}$ and Ni-$d_{z2}$ orbitals, respectively.

*Atomic and electronic structure.* To discuss the atomic and electronic properties of interface contain apical oxygen, we focus on sandwich structure (STO)$_4$|(NdNiO$_2$)$_{10}$|(STO)$_1$ with single apical oxygen at interfacial Nd layer as shown in Fig. 5(a) (For simplicity, we used single oxygen in supercell instead of 0.75 oxygen in DFT calculation, because they have similar atomic and electronic structure). This structure can be regarded as half perovskite NdNiO$_3$ at the interface and NdNiO$_2$ in subsequent layers. The appearance of apical oxygen at the interface will change the crystal structure. For the apical oxygen-free structure, the distances between neighboring Nd layers increase continuously from ~3.216 Å at the NiO$_2$/Nd/TiO$_2$ interface to ~3.268 Å at the center (Fig. 1). Whereas an opposite trend is observed for the apical oxygen structure, in which the distances decrease as moving from the interface ~3.358 Å to the center ~3.235 Å (Fig. 5(a)). This is because of the additional oxygen atom and Ni-O bond at the interface Ni layer, leading to an enhancement of the out-of-plane lattice constant at the interface. In apical oxygen-free structure, the central NiO$_2$ layers has coplanar structure as well as that in bulk, whereas the interfacial layer is buckled, and the Ni is moved outwards perpendicularly from its NiO$_2$ layer by $d_z$ = −0.186 Å at NiO$_2$/Nd/TiO$_2$ interface and $d_z$ = 0.225 Å at Nd/NiO$_2$/SrO$_2$ interface, and Ti at the NiO$_2$/Nd/TiO$_2$ interface vertically displaces inwards considerably. This is also different from that of apical oxygen structure: $d_z$ = 0.182 Å at NiO$_2$/NdO/TiO$_2$ interface and $d_z$ = 0.215 Å at Nd/NiO$_2$/SrO$_2$ interface, and interfacial Ti shows a negligible displacement. Measuring these different structural responses could be an indirect way of detecting the apical oxygen at the interface in transmission electron microscopy (TEM) under the condition that the TEM resolution is not enough to observe small oxygen atoms directly[55].

The apical oxygen at the interface leads to a substantial change of the electronic structure. We can observe from the calculated PDOS in Fig. 5(a) that the interfacial STO conduction band across the Fermi level is nearly zero, suggesting that the STO layers is insulating. Fig. 5(b, c) shows the orbital-decomposed PDOS of central bulk-like layer and interfacial layer. It reveals that in the central bulk-like layer, the main contribution of Ni-*3d* near the Fermi level is almost from the $d_{x^2-y^2}$ state. Whereas in the interfacial layer, Ni is surrounded by apical oxygen compared to the bulk-like NdNiO$_2$ layer, and the oxygen will tend to take extra electron from Ni, so that the $d_{z^2}$ contribution becomes more pronounced[26]. Therefore, it exhibits a mixture of $d_{x^2-y^2}$ and $d_{z^2}$ states near the Fermi level at the interface layer. Such a change of electronic structure also leads to the magnetic moment of interface Ni increases to ~1.5 μ$_B$, whereas the magnetic moment of central Ni is smaller than 1.0 $\mu_B$, which is the same as the bulk one. The orbital projected electronic band structures of NdNiO$_2$ are shown in Fig. 5(d, e). For the central Ni, we observe a single-band feature of $d_{x^2-y^2}$ state near Fermi level, and that $d_{z^2}$ state locates far away from Fermi level, which is similar to the bulk one[5]. For the interfacial Ni, however, $d_{z^2}$ state locates quite close to the Fermi level and becomes more pronounced because of the charge transfer from the apical oxygen to Ni, which leads to $d_{z^2}$ and $d_{x^2-y^2}$ states hybrids at Γ-X direction, and overlaps at M-Γ direction. Such a band hybrid weakens the single-band feature and is thought to be detrimental to superconductivity[56]. Therefore, we can conclude that superconductivity in NdNiO$_2$ is likely not originated from the interface.

## IV. CONCLUSION

To summarize, using first-principles DFT calculations and charge transfer self-consistent model, we demonstrated that the polar instability will lead to atomic reconstruction at NdNiO$_2$/STO interface. We further proposed a reasonable rough interface configuration with residual apical oxygen in the interfacial Nd layer, which could screen the built-in electric field and avoid polar instability. Such proposals can be tested by future TEM experiments. The appearance of apical oxygen destroys the single-band electronic structure at the interface, likely rendering the interface irrelevant for the experimentally observed superconductivity. Thus, our work indicates that the polar discontinuity in infinite-layer nickelate superconductor may suppress the emergence of superconductivity at the interface. Further consideration for such limitation needs to be taken in the structural design of nickelates superconductors as well as other oxide heterostructures.

**ACKNOWLEDGMENTS**
This work was supported by the National Key R&D Program of China (Grant No. 2017YFA0303602), the National Nature Science Foundation of China (Grants No. 11774360, No. 11974365, and No. 51931011), and Key Research Program of Frontier Sciences, CAS (No. ZDBS-LY-SLH008). Y.L. acknowledges support by Deutsche Forschungsgemeinschaft (DFG) under Germany's Excellence Strategy EXC2181/1-390900948 (the Heidelberg STRUCTURES

Excellence Cluster). M.C. was supported by the National Natural Science Foundation of China (Grant No. 11774084) and the Project of Educational Commission of Hunan Province of China, 18A003. Calculations were performed at the Supercomputing Center of Ningbo Institute of Materials Technology and Engineering.